\documentclass[]{spie}

\usepackage{amsmath,amsfonts,amssymb}
\usepackage{graphicx}
\usepackage[colorlinks=true, allcolors=blue]{hyperref}
\usepackage{caption}
\usepackage{subcaption}

\title{Thermal architecture for the QUBIC cryogenic receiver}

\author[t]{A.J. May}
\author[a]{C. Chapron}
\author[t]{G. Coppi}
\author[z,m]{G. D'Alessandro}
\author[z,m]{P. de Bernardis}
\author[z,m]{S. Masi}
\author[t]{S. Melhuish}
\author[a]{M. Piat}
\author[t]{L. Piccirillo}
\author[z,c]{A. Schillaci}
% ...

\affil[a]{Astroparticule et Cosmologie (CNRS-IN2P3), Paris, France}
% ...
\affil[c]{California Institute of Technology, Pasadena, CA, USA}
% ...
\affil[m]{INFN, Sezione di Roma 1	, Rome, Italy}
% ...
\affil[t]{Jodrell Bank Centre for Astrophysics, University of Manchester, Manchester, UK}
% ...
\affil[z]{Universit\`{a} di Roma - La Sapienza, Rome, Italy}
% ...

\authorinfo{Further author information:\\A.J.M.: E-mail: andrew.may-3@postgrad.manchester.ac.uk, Telephone: +44 (0) 161 306 6470\\  S.M.: E-mail: silvia.masi@roma1.infn.it}

\begin{document} 
\maketitle

\begin{abstract}

QUBIC, the QU Bolometric Interferometer for Cosmology, is a novel forthcoming instrument to measure the B-mode polarization anisotropy of the Cosmic Microwave Background. The detection of the B-mode signal will be extremely challenging; QUBIC has been designed to address this with a novel approach, namely bolometric interferometry. The receiver cryostat is exceptionally large and cools complex optical and detector stages to 40~K, 4~K, 1~K and 350~mK using two pulse tube coolers, a novel $^{4}$He sorption cooler and a double-stage $^{3}$He/$^{4}$He sorption cooler. We discuss the thermal and mechanical design of the cryostat, modelling and thermal analysis, and laboratory cryogenic testing.
\end{abstract}


\section{INTRODUCTION}
\label{sec:intro}  

QUBIC, the QU Bolometric Interferometer for Cosmology\cite{tartari2016qubic,aumont2016qubic,osullivan}, is a novel forthcoming instrument to measure the B-mode polarization anisotropy of the Cosmic Microwave Background at intermediate angular scales (30~$<$~$\ell$~$<$~200). The detection of primordial B-mode polarization would constitute direct evidence for inflation \cite{seljak1997signature,kamionkowski1997probe}. Measurement of the signal will be extremely challenging, owing primarily to the size of the expected signal, instrumental systematics that could possibly induce polarization leakage from the relatively large E signal into B, and the brightness of foregrounds (principally polarized galactic dust emission)\cite{aumont2016qubic}. QUBIC has been designed to address this challenge with a novel approach, namely bolometric interferometry, which combines the background-limited sensitivity of transition edge sensors (TES) with the control of systematics allowed by the observation of interference fringe patterns.

The instrument will directly observe the sky through a horn array, the signals from which will be combined in an optical combiner. Polarization modulation will be achieved using a cold stepped half-wave plate (HWP). Images of the interference fringes will be formed on two focal planes tiled with TES bolometers allowing operation at two frequency bands to support foreground disentanglement. Deployment of the first QUBIC module at the Alto Chorrillos site in Argentina is planned in two stages: the first with a reduced number of detectors operating at 150 GHz before the end of 2018 (referred to as the technological demonstrator, or TD) and then an upgrade to the full number of detectors with both 150 and 220 GHz channels in 2019 (final instrument, FI).

Given the size and complexity of the receiver optical chain (detailed in the Section~\ref{sec:instrument}), cryogenic design has been exceptionally challenging. Extensive modelling and thermal analysis, as well as laboratory cryogenic testing of various subsystems, have been required to support the thermal and mechanical design of the cryostat. Additionally, in view of the size of the 1~K box and level of thermal isolation, significant efforts have been devoted to developing a precooling strategy to support reasonable cooldown times for both laboratory testing of the receiver and deployment. 

\section{INSTRUMENT ARCHITECTURE}
\label{sec:instrument}

A detailed schematic of the QUBIC instrument architecture is shown in Figure~\ref{fig:instrument} below; note the operating temperatures of each element in the receiver chain.

\begin{figure} [ht]
\begin{center}
\begin{tabular}{c} 
\includegraphics[height=9cm]{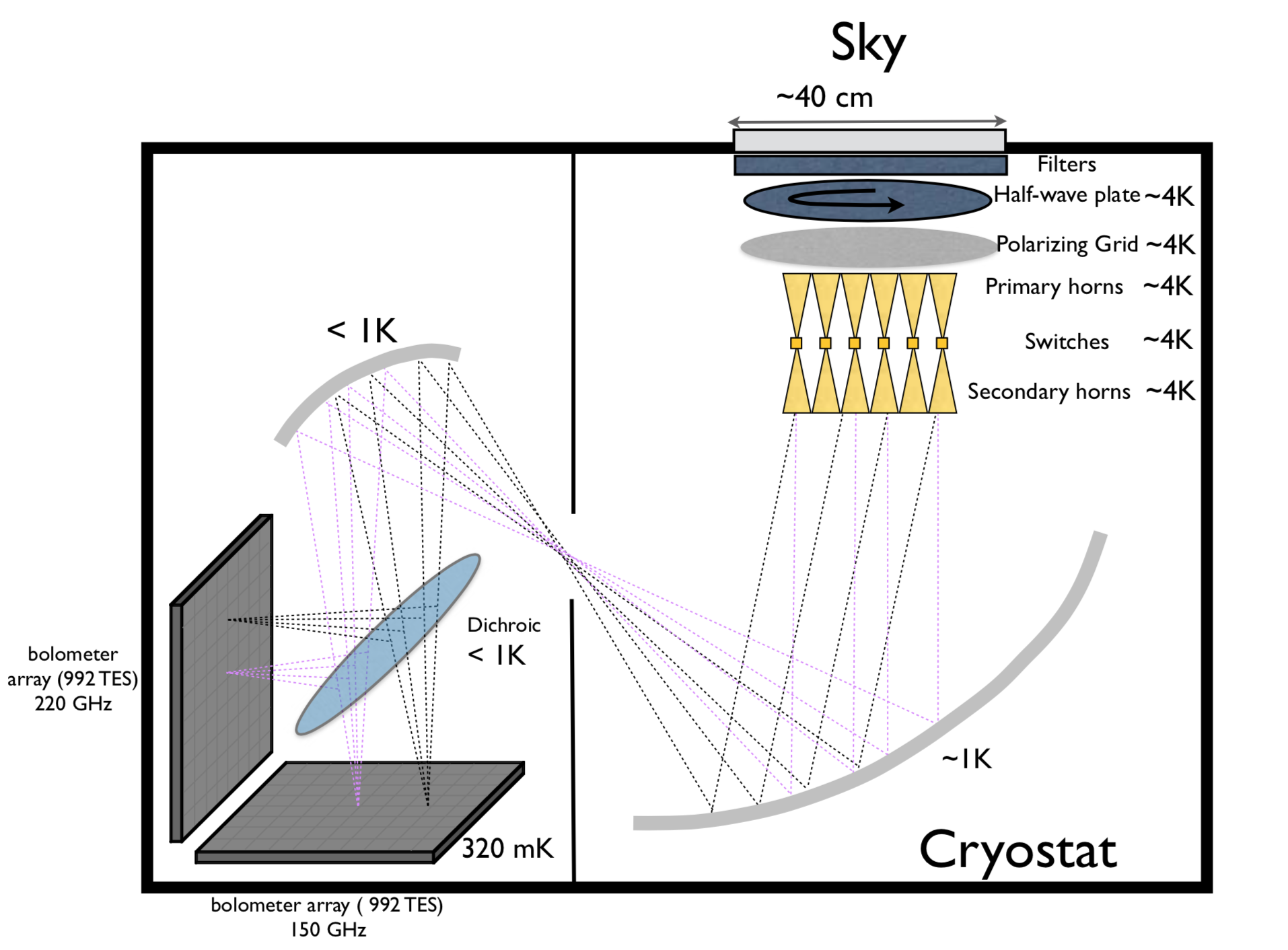}
\end{tabular}
\caption{QUBIC instrument schematic} 
\label{fig:instrument}
\end{center}
\end{figure}

The signal from the sky enters the cryostat vacuum through a 40 cm clear aperture window, fabricated from ultra high molecular weight (UHMW) polyethylene (PE)\cite{d2018ultra}. A series of single-layer metal-mesh element low-pass filters are stacked at 300~K, 40~K and 4~K to minimise radiative thermal loading on the cold stages by sequentially rejecting short wavelength radiation. 

The polarization is then modulated by a cold stepped HWP. The HWP used is a broadband metal mesh design\cite{pisano2012broadband}, 30 cm in diameter, and is operated at 4~K. One polarization state is then selected by a photolithographic polarizing grid, also at 4~K.

An array of 400 back-to-back corrugated horns at 4~K then collects and re-images the radiation\cite{scully2016optical}. The horn array is composed of two blocks of 400 horns each, placed back-to-back with interstitial switches that can open or close the optical path to the radiation; this allows for the self-calibration technique which is based on comparing redundant baselines produced by equally spaced pairs of horns\cite{bigot2013self}.

The radiation is reimaged onto a dual-mirror optical combiner (at 1~K)\cite{scully2016optical} that sums all the signals picked-up by the sky-side horns onto the detector stage. 

A dichroic filter (1~K) placed between the optical combiner and the focal planes selects the two frequency bands, centered at 150~GHz and 220~GHz, which are focused onto two orthogonal focal planes. The focal planes, operating at 350~mK, are each tiled with 1024 transition edge sensor (TES) detectors with a critical normal-to-superconducting temperature close to 500 mK. The TESs are read out by a time domain multiplexing (TDM) scheme\cite{salatino} based on superconducting quantum interference devices (SQUIDs) operating at 1~K. The signals from the SQUIDs are then amplified by SiGe low noise amplifier (LNA) based application-specific integrated circuits (ASICs) at 40~K and read out by warm electronics at 300~K.

The instrument design, based on this architecture, has been finalised\cite{aumont2016qubic}. At the time of writing, all of the subsystems have been validated\cite{osullivan} and have been shipped to the Laboratoire Astroparticule et Cosmologie (APC) for integration and testing before deployment to the Alto Chorrillos site later this year. Figure~\ref{fig:fullCAD} below shows a section-view of the full receiver cryostat model. The cryostat measures 1.4 m in diameter and 1.55 m in height; when the forebaffle cone is mounted around the window, the total height will be 1.9 m and the total instrument mass will be 800 kg.

\begin{figure} [ht]
\begin{center}
\begin{tabular}{c} 
\includegraphics[height=10cm]{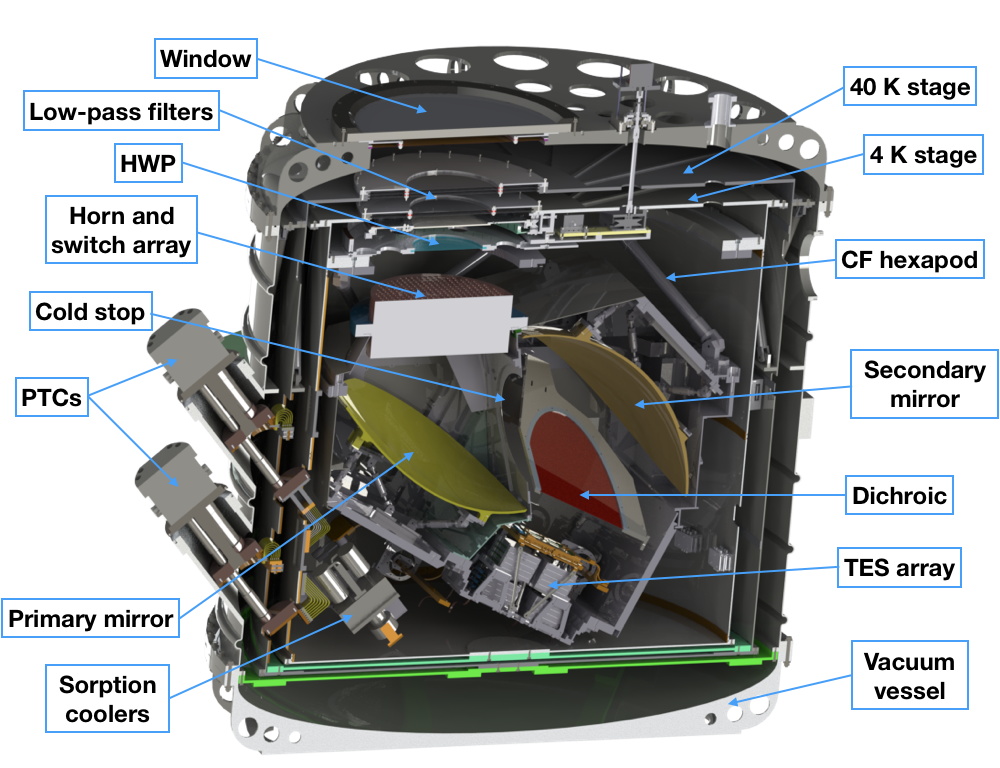}
\end{tabular}
\caption{QUBIC receiver CAD model; the cryostat measures 1.4 m in diameter and 1.55 m in height} 
\label{fig:fullCAD}
\end{center}
\end{figure}

\section{VACUUM CHAMBER, 40~K AND 4~K STAGES}
\label{sec:4K}

Given the thermal requirements for the subsystems outlined in Section~\ref{sec:instrument}, the cryostat design is based on relatively low mass 40~K and 4~K stages with a large and complex 1~K stage and a subsequent very low mass 350~mK stage (note that all temperature stages are referred to by their nominal operating temperatures). The following subsections describe the vacuum chamber, 40~K and 4~K stages; the 1~K and 350~mK stages are described in Sections~\ref{sec:1K} and~\ref{sec:350mK}.

\subsection{VACUUM CHAMBER}

In order to house all of the optical elements described in Section~\ref{sec:instrument} (see especially Figures~\ref{fig:instrument} and~\ref{fig:fullCAD}), a large (order of 1 m$^{3}$) vacuum chamber is required.

The shape and structure of the jacket have been optimised for withstanding the stress of atmospheric pressure outside the cryostat. The structure is a vertical cylinder with top and bottom bulkheads, fabricated from aluminium alloy sheets, as shown in Figure~\ref{fig:Picture3} below. The cylinder is roll-bent and welded, with reinforcing ribs. The shell has diameter of 1.4 m and height of 1.55 m. Buna-N o-ring seals are used to hold the vacuum.

The optical window on the top bulkhead (see Figure~\ref{fig:fullCAD}) is fabricated from high-density PE. A cylindrical slab (560 mm diameter, 20 mm thick) of PE has been used to meet the structural requirement whilst providing excellent mm-wave transmission\cite{d2018ultra}.

\subsection{40~K and 4~K stages}

The first cold stage is a 40~K shield which intercepts radiative loading from the vacuum chamber at 300~K. This stage also supports low-pass filters as described in Section~\ref{sec:instrument}. The shield is fabricated from aluminium and is wrapped in superinsulation to reduce the effective emmissivity\cite{bapat1990performance}. Copper strips are bolted to the aluminium to help improve the uniformity of the temperature distribution in the shields. Fibreglass tubes arranged in a hexapod configuration are used to mount the shield off the vacuum chamber. Figure~\ref{fig:Picture1} below shows the 40~K shield being mounted.

The next temperature stage is a 4~K shield, which is of a similar construction and mounted concentrically inside the 40~K shield. A second hexapod is used to mount the 4~K shield off the 40~K stage to minimise parasitic conductive loading.

Cooling of the 40~K and 4~K stages is provided by two Sumitomo RP-082B2S\footnote{www.shicryogenics.com} pulse tube cryocoolers (PTCs) operating in parallel. Each cooler has a nominal capacity of 0.9 W at 4.2 K and 35 W at 45 K. Copper straps between the PTC heads and the shields allow for differential thermal contraction during cooldown.

Thermal models of the cryostat give an estimated loading of 0.1 W on the 4~K stage and 16 W on the 40~K stage; as such, operation with a single PTC is possible. However, two PTCs are used in order to both reduce the cooldown time (see below and Section~\ref{subsec:precool}) and to accept the large intermittent loading associated with recycling the 1~K and 350~mK fridges (see Sections~\ref{sec:1K} and~\ref{sec:350mK}). 

As shown in Figures~\ref{fig:fullCAD} and~\ref{fig:Picture3}, the longitudinal axes of the two PTCs are tilted by 40$^{\circ}$ with respect to the vertical; this allows observations with elevations between 30$^{\circ}$ and 70$^{\circ}$ whilst maintaining the PTC heads within 20$^{\circ}$ of vertical where cooling performance is maximised.

\begin{figure}[h]
\centering
\begin{minipage}{.45\textwidth}
  \centering
  \includegraphics[height=9cm]{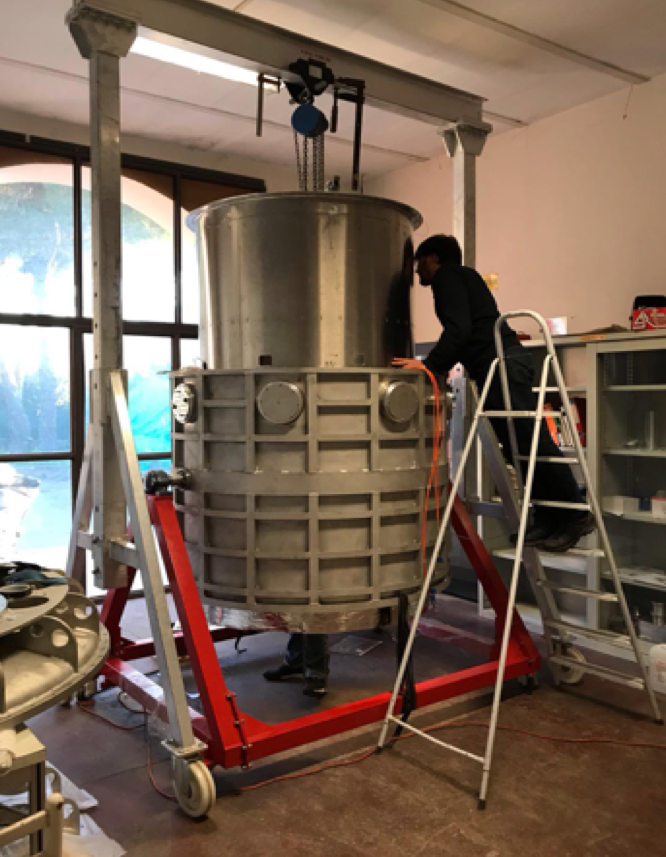}
  \captionof{figure}{40~K shield mounting in cryostat}
  \label{fig:Picture1}
\end{minipage}%
\begin{minipage}{.45\textwidth}
  \centering
  \includegraphics[height=9cm]{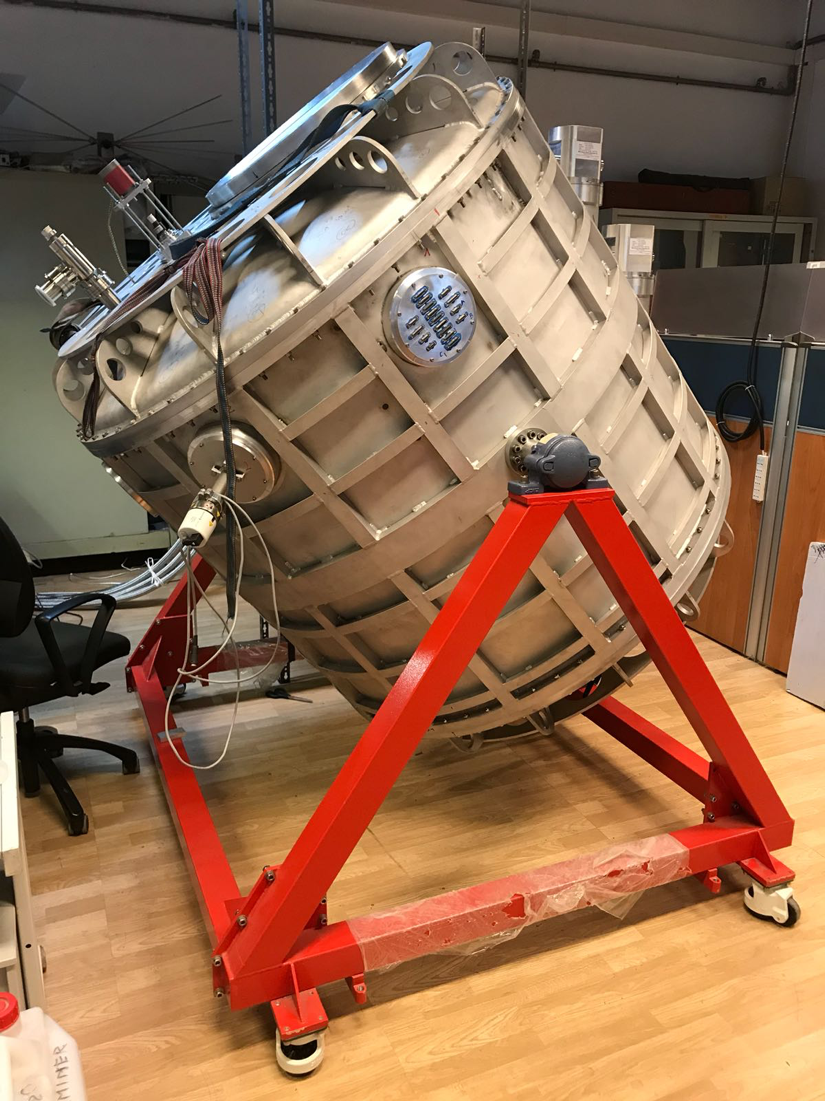}
  \captionof{figure}{Closed cryostat under test}
  \label{fig:Picture3}
\end{minipage}
\end{figure}

\begin{figure}[ht]
\begin{center}
\begin{tabular}{c} 
\includegraphics[height=7cm]{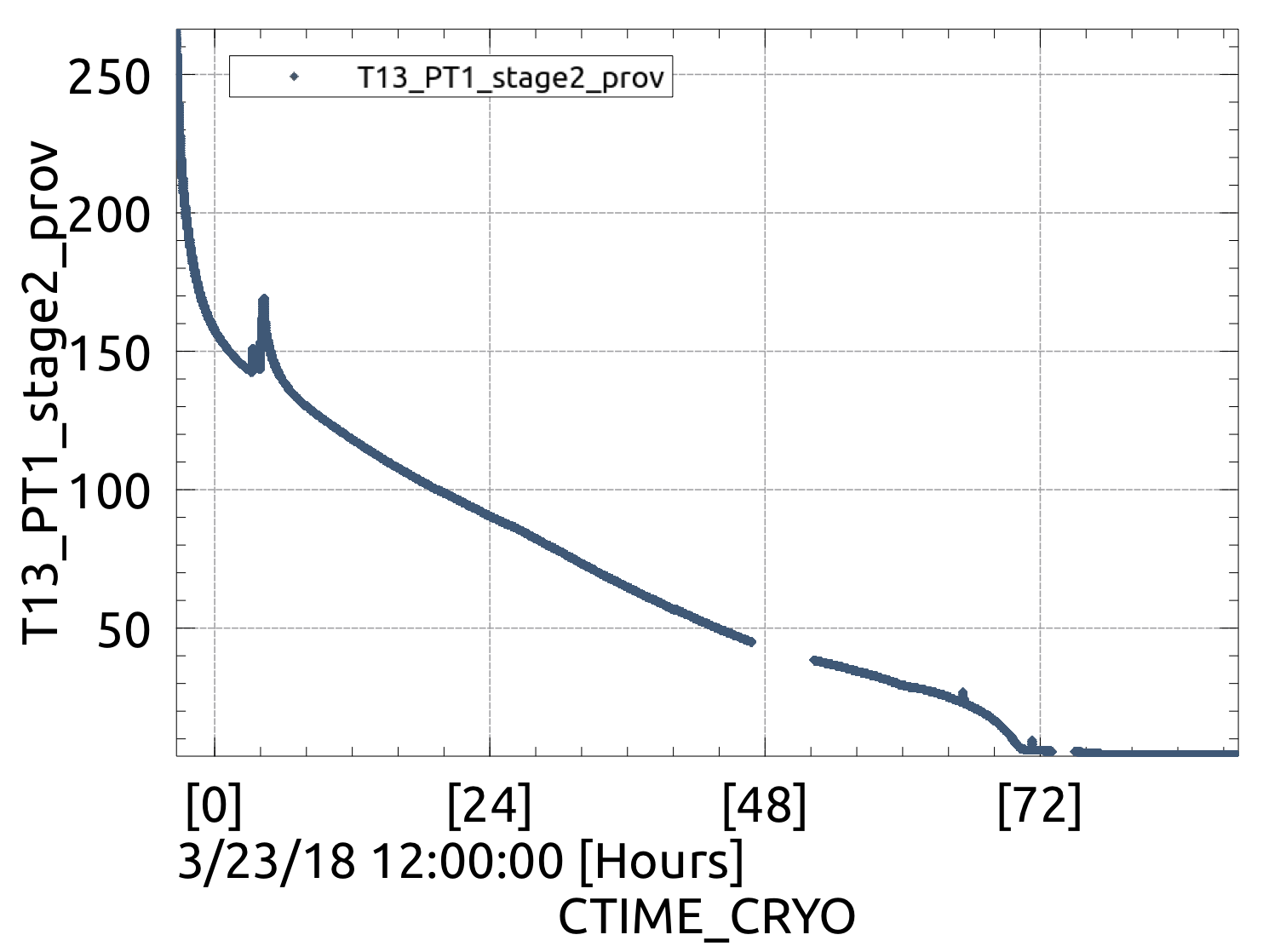}
\end{tabular}
\caption{Cooldown of the 4~K stage during test in Roma; spikes are due to periods where the two PTCs have been switched off to switch from mains power to UPS} 
\label{fig:cooldown}
\end{center}
\end{figure}

Initial cooldown testing of the cryostat with the 40~K and 4~K shields mounted has been carried out in Roma. This testing was done to validate the integrity of the vacuum jacket and the performance of the PTCs; both have been successful. The measured cooldown time was 72 hours, with a final temperature of the 4~K stage of 3.1$\pm$0.5 K, as shown in Figure~\ref{fig:cooldown}.

As well as the radiation shield, the 4~K stage also supports further low-pass filters, the horn array and HWP. In order to step the HWP, a cryogenic rotation mechanism is needed. The mechanism designed for QUBIC inherits several of the solutions developed for the PILOT balloon-borne instrument successfully flown by CNES in 2015 and 2017\cite{salatino2011cryogenic}. The HWP is rotated by a stepper motor mounted outside the vacuum chamber (at 300~K). Rotation is transmitted through the jacket by means of a magnetic joint to a fibreglass axle. This axle then runs down to the 4~K stage and drives a Kevlar belt which is tensioned by a spring-loaded capstan pulley. The HWP is mounted inside a support ring which has a groove for the Kevlar belt. The mounting of the HWP uses a custom block in order to reduce the differential thermal contraction between the support ring and the HWP. An optical encoder records the position of the ring. 

\begin{figure}[h]
\centering
\begin{minipage}{.5\textwidth}
  \centering
  \includegraphics[height=5.5cm]{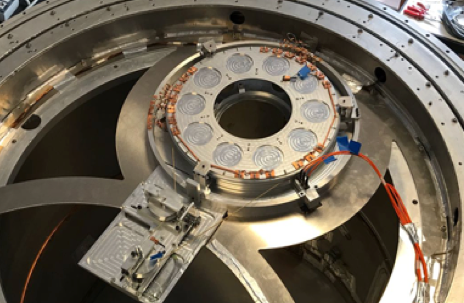}
  \captionof{figure}{Half wave plate rotator mounted in cryostat}
  \label{fig:Picture5}
\end{minipage}%
\begin{minipage}{.5\textwidth}
  \centering
  \includegraphics[height=5.5cm]{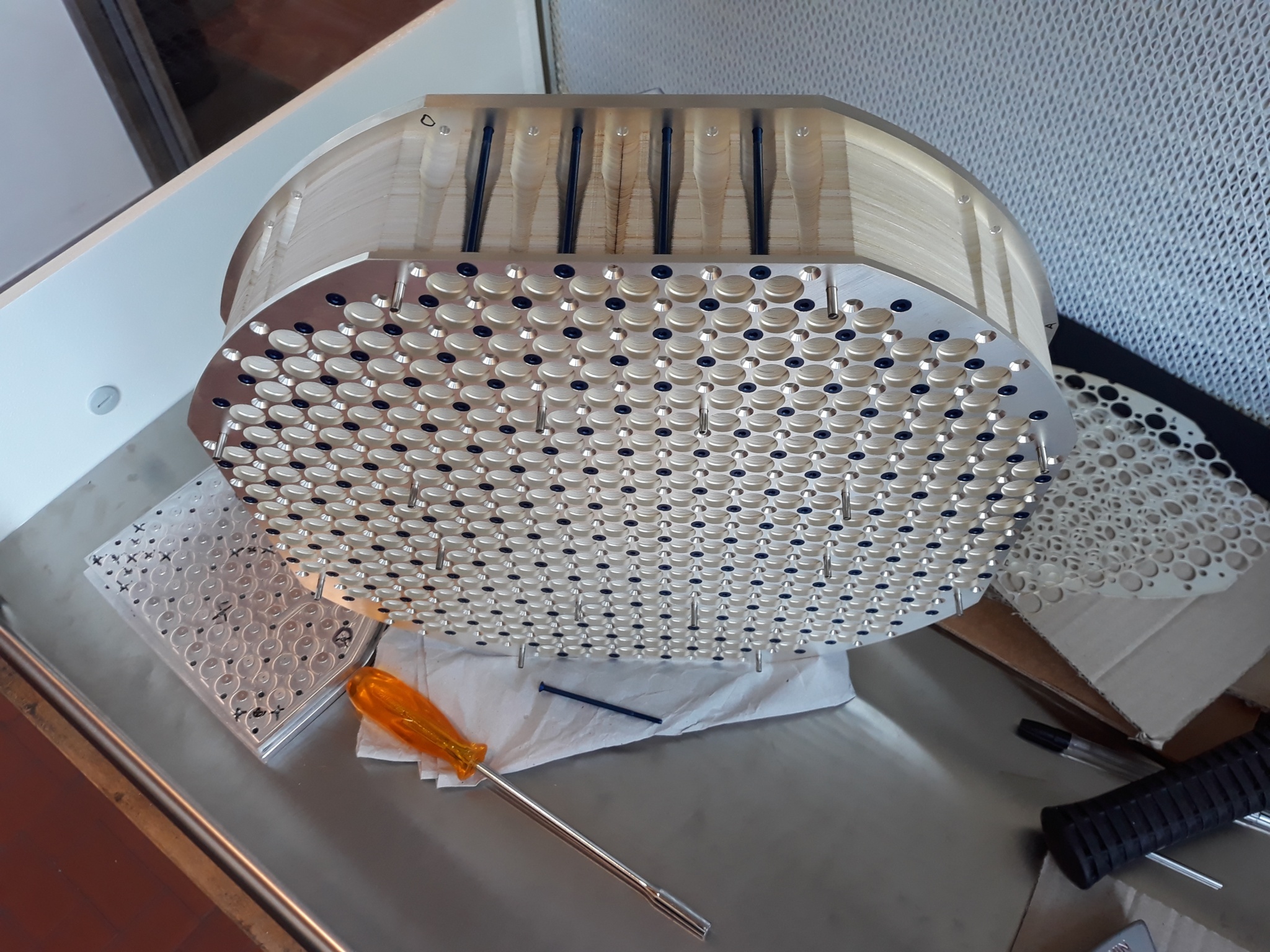}
  \captionof{figure}{One block of 400 horn module}
  \label{fig:20180424_101331}
\end{minipage}
\end{figure}

The rotator is shown under test in the cryostat in Roma in Figure~\ref{fig:Picture5}. One block of the horn array is shown during assembly in Milano in Figure~\ref{fig:20180424_101331}.

Having completed validation testing, both the cryostat and these subsystems have been shipped to APC for integration and testing.

\section{1~K STAGE}
\label{sec:1K}

Given the requirements for a number of optical elements to be operated at 1~K (see Section~\ref{sec:instrument}), a 1~K stage is mounted off the 4~K stage described above. This stage includes the 1~K optics box, precooling heat switches for the box, a $^{4}$He 1~K cooler and an isolation switch for the cooler; these subsystems are described in Sections~\ref{subsec:1Kbox},~\ref{subsec:precool},~\ref{subsec:1Kcooler} and~\ref{subsec:isoswitch} respectively.

\subsection{1~K BOX}
\label{subsec:1Kbox}

The 1~K optics box contains the primary and secondary mirrors (M1 and M2) for the optical combiner, the cold stop, dichroic and focal plane assembly. The CAD model is shown in Figures~\ref{fig:1KboxCAD2} and~\ref{fig:1KboxCAD} below.

\begin{figure}[h]
\centering
\begin{minipage}{.5\textwidth}
  \centering
    \includegraphics[height=6.2cm]{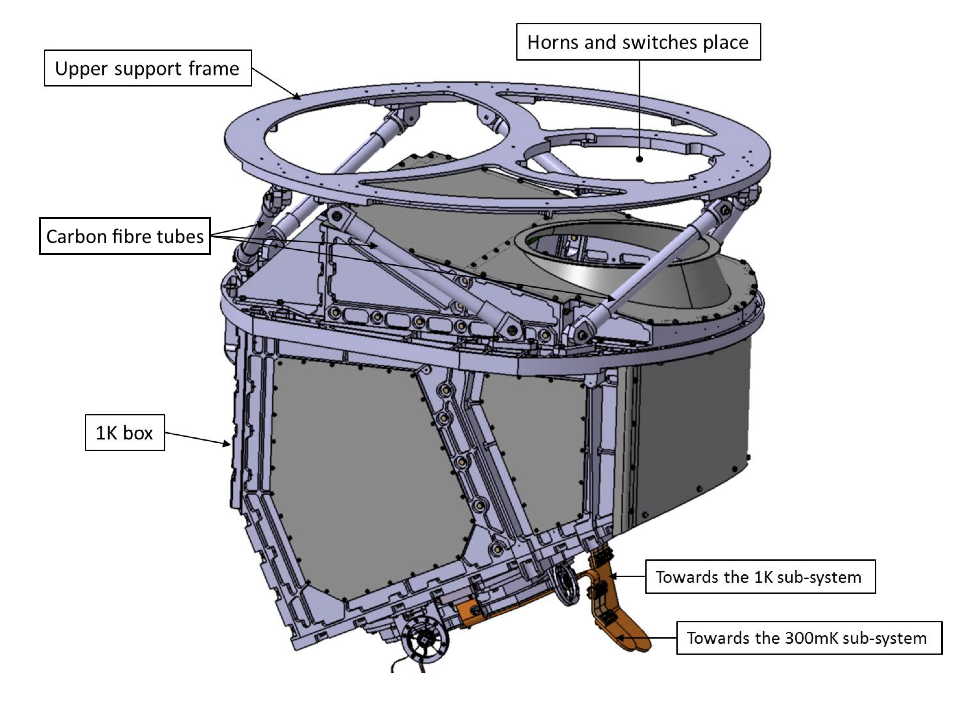}
  \captionof{figure}{1~K optics box CAD}
  \label{fig:1KboxCAD2}
\end{minipage}%
\begin{minipage}{.5\textwidth}
  \centering
\includegraphics[height=6.2cm]{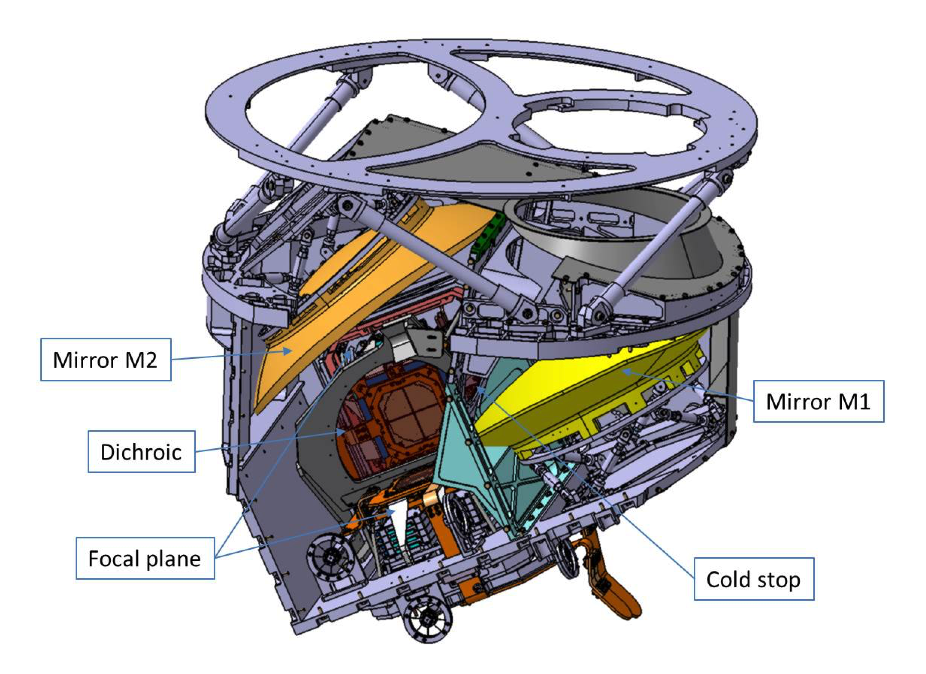}
  \captionof{figure}{1~K optics box CAD (section view)}
  \label{fig:1KboxCAD}
\end{minipage}
\end{figure}

The mirrors are designed at their final temperature and machined in aluminium so as to reach their desired geometries after thermal contraction down to 1~K. The purpose of the optics box is to provide both the required thermal environment and mechanical support/alignment for the optics. Furthermore, it provides thermal shielding at 1~K for the 350~mK stage (see Section~\ref{sec:350mK}). 

The box contains 140~kg of aluminium 6061, 11~kg of stainless steel 304, 10~kg of copper and 4~kg of brass, giving a total mass of $\sim$165~kg. Given these masses and the specific heats of these materials\cite{pobell2007matter}, $\sim$20~MJ needs to be removed to cool the box from 300~K to 1~K.

As such, significant efforts have been devoted to optimising the cooldown of this stage (see Section~\ref{subsec:precool}). The solution that has been chosen is to use heat switches between the 1~K box and 4~K stage to utilise the high cooling power of the PTCs (see Section~\ref{sec:4K}) in order to precool the box to 4~K; these switches are then opened before the box is cooled further to 1~K by the $^{4}$He sorption cooler described in Section~\ref{subsec:1Kcooler}. In order to reduce thermal impedances between components within the 1~K box (and hence reduce the cooldown time), a thin layer of Apiezon N grease\footnote{www.apiezon.com} will be applied between all of the bolted interfaces.

\begin{figure} [ht]
\begin{center}
\begin{tabular}{c} 
\includegraphics[height=6cm]{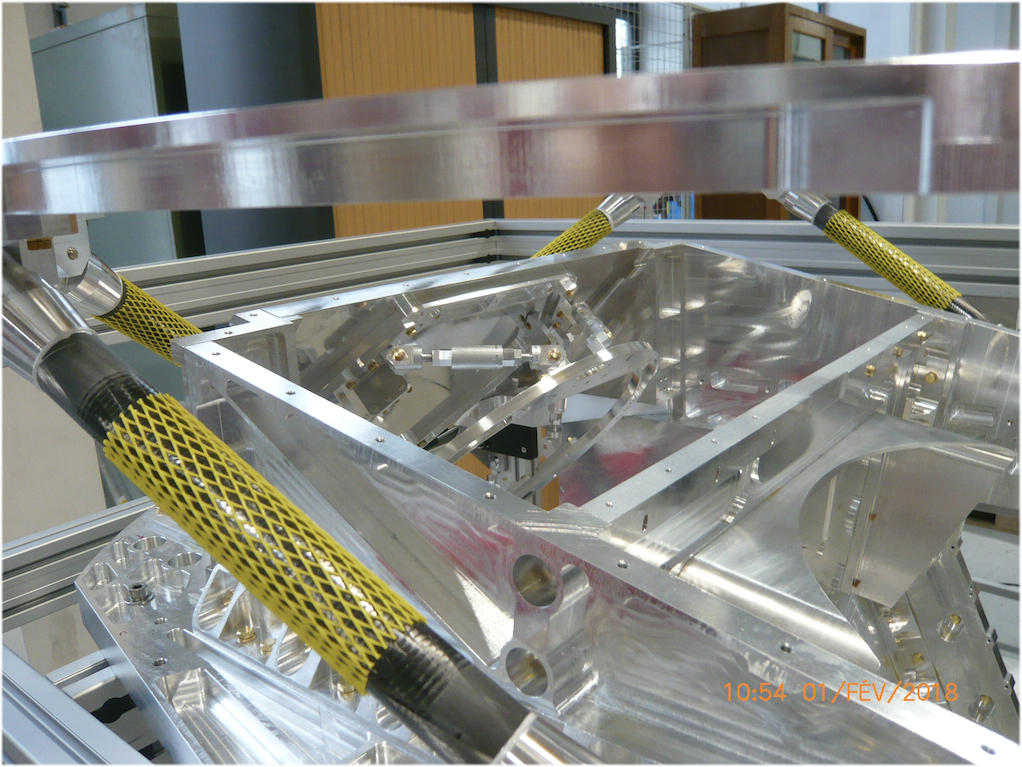}
\end{tabular}
\caption{Carbon fibre tie rod hexapod assembly for 1~K box} 
\label{fig:GDA2}
\end{center}
\end{figure}

The 1~K box is mounted off the 4~K stage as shown in Figure~\ref{fig:1KboxCAD2}. A hexapod of carbon fibre tie rods is used to meet the mechanical requirements whilst minimising parasitic conductive loading. An image of the assembled rods is shown in Figure~\ref{fig:GDA2}. In order to validate the thermal and mechanical performance of the tie rods, a number of tests were carried out at APC and CEA Saclay. The measured properties are given in Table~\ref{tab:matprops1}. These measurements were used to calculate the loading through the hexapod and are reported along with additional sources of parasitic loading in Table~\ref{tab:1K}. This loading gives the steady state loading requirement for the 1~K cooler described in Section~\ref{subsec:1Kcooler}.

\begin{table}[h]
\centering
\begin{minipage}{.48\textwidth}
  \centering
    \caption{Measured carbon fibre tube properties}
      \label{tab:matprops1}
    \begin{tabular}{| l | l |}
  \hline			
  Parameter & Value \\
  \hline
  Fibre content by volume & 60 \% \\
  DP406 resin content by volume & 40 \% \\
  External diameter & 32 mm \\
Internal diameter & 30 mm \\
Young's modulus (at 300~K) & 120 GPa \\
Young's modulus (at 4.2 K) & 80 GPa \\
Conductance integral 5 K to 1~K & 12.7 mW/m \\
  Axial expansion 80 K to 280 K & -0.014 \% \\
  Radial expansion 80 K to 280 K & 0.12 \% \\
  \hline  
\end{tabular}
\end{minipage}%
\begin{minipage}{.51\textwidth}
  \centering
    \caption{1~K box steady-state thermal loading}
  \label{tab:1K}
    \begin{tabular}{| l | l |}
  \hline			
  Source & Value \\
  \hline
  6 tie rods (epoxy/carbon-fibre) & 0.32 mW \\
  2 precooling switches (see Section~\ref{subsec:precool}) & 0.33 mW \\
  Instrumentation wires & 15 $\mu$W \\
  Thermal radiation through window & 1 $\mu$W \\
  Thermal radiation ($\varepsilon$=0.1) 4.2 K to 1 K & 6 $\mu$W \\
  Total & 0.7 mW \\
\hline   
\end{tabular}
\end{minipage}
\end{table}

\subsection{PRECOOLING HEAT SWITCHES}
\label{subsec:precool}

Given the thermal isolation of the 1~K box that is required to meet the steady-state heat load requirements (see Section~\ref{subsec:1Kbox}), first-order transient thermal simulations have shown cooldown times on the order of a week. Furthermore, given the assumptions in these models, it is possible that the cooldown time could be significantly longer. For both development testing and operation after deployment, it is highly desirable to reduce the cooldown time appreciably. 

Previous experiments\cite{Lounasmaa1974} have achieved faster cooldown of mechanically well isolated sub-4~K stages through the use of a low pressure exchange gas in the vacuum space to couple the isolated stage to a 4~K stage for precooling. The sub-4~K stage can then be isolated by removing the exchange gas before cooling to its base temperature. This strategy is unsuitable for QUBIC however, due to both the potential damage to optical components due to differential thermal contraction and the well-known affinity of helium to accumulate in the interstitial space of MLI \cite{bapat1990performance}. As MLI will be used on both the 40~K and 4~K stages of the receiver, the use of exchange gas in this case would accordingly require significant pumping time to remove after precooling and would therefore be expected to increase the total cooldown time significantly. 

An alternative and far more attractive strategy is the use of heat switches to couple to 1~K box to the 4~K stage. In order to support this, switches must be used that are reliable and have high switching ratios (i.e., the ratio between closed and open conductance) such that the precooling time to 4~K is minimised whilst at the same time not significantly increasing the steady-state heat leak at 1~K.

Based on open and closed conductance measurements of several types of heat switch\cite{aumont2016qubic,May4}, it has been decided to employ a combination of a mechanical switch (conductance on the order of 0.5 W/K above $\sim$60 K but falls off sharply below this temperature) and two convective heat switches (higher conductance than the mechanical switch below $\sim$40~K, relatively constant with temperature down to 4~K).

Mechanical heat switches generally operate by actuating some mechanism in order to make or break a physical connection. In this way, zero open conductance is possible and very high closed conductances may be achieved as long as the required pressure may be developed at the interface. A mechanical switch (available commercially from Entropy\footnote{http://www.entropy-cryogenics.com}) will be used; the switch is shown under test in a cryostat at APC in Figure~\ref{fig:mechhs} below.

\begin{figure} [ht]
\begin{center}
\begin{tabular}{c} 
\includegraphics[height=6cm]{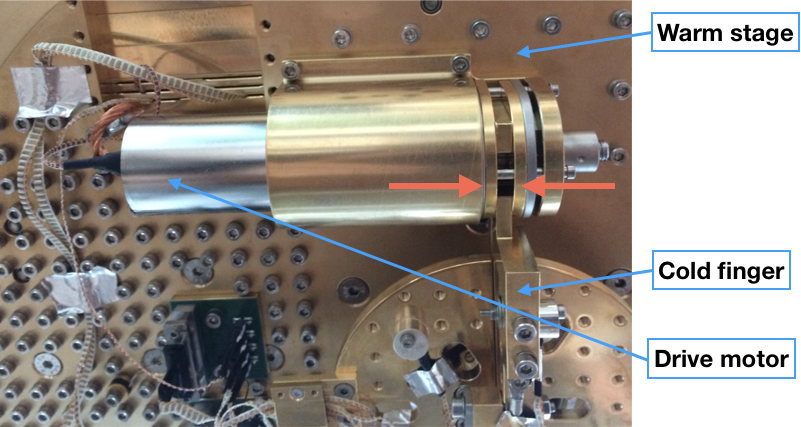}
\end{tabular}
\caption{Entropy mechanical heat switch under test at APC. The red arrows indicate how the switch (mounted on the 4~K stage) operates to clamp with high pressure to the cold finger (1~K). The length of the switch is 131 mm.} 
\label{fig:mechhs}
\end{center}
\end{figure}

\begin{figure}[!ht]
\centering
\begin{minipage}{.35\textwidth}
  \centering
  \includegraphics[width=\linewidth]{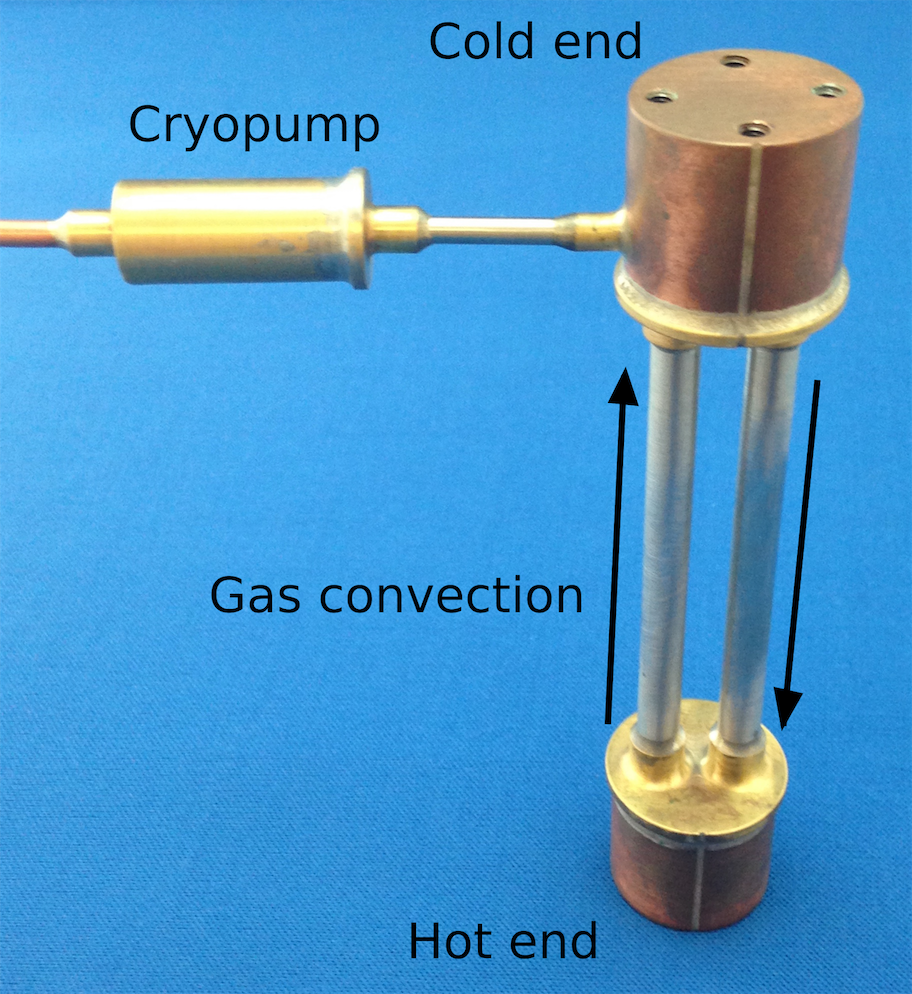}
  \caption{Photograph of convective switch; the height of the switch shown here is 131 mm}
    \label{fig:convHS}
\end{minipage}%
\begin{minipage}{.6\textwidth}
  \centering
  \includegraphics[width=0.9\linewidth]{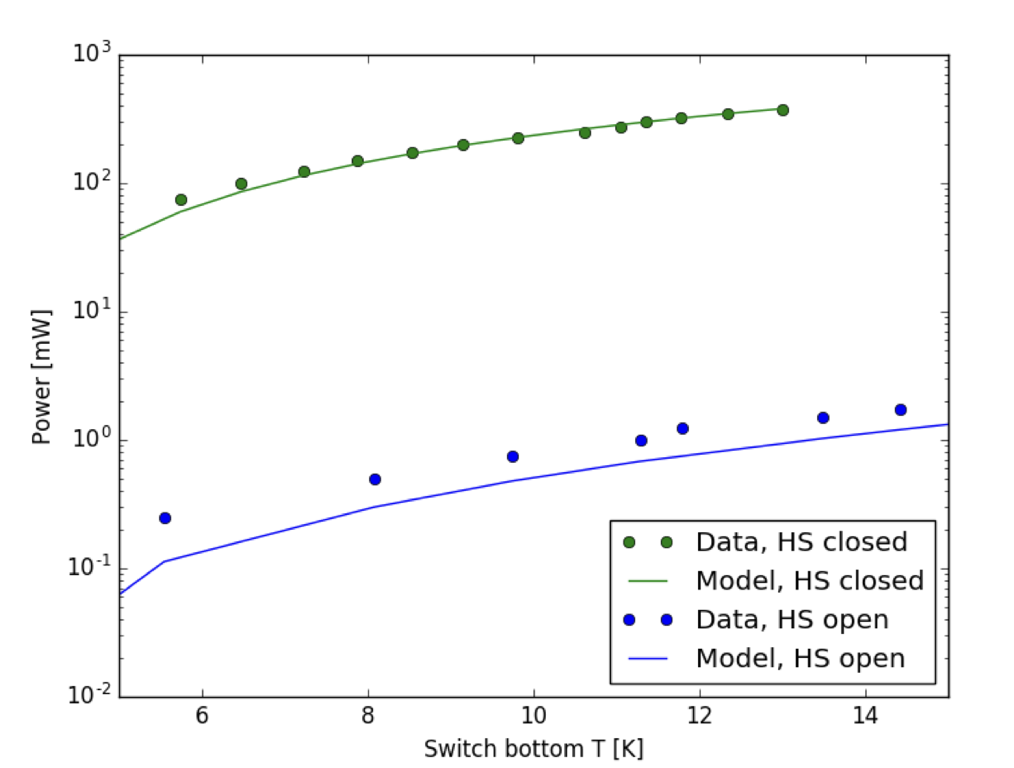}
\caption{Open and closed conductance of convective switch}
\label{fig:precoolHScond1}
\end{minipage}
\end{figure}

To provide a high conductance connection below 60 K, two convective heat switches\cite{piccirillo2018miniature,May4} are used in parallel with the mechanical switch. Each convective switch essentially consists of a circuit comprising two stainless steel tubes and a copper heat exchanger at either end connected to the warm and cold stages (as shown in Figure~\ref{fig:convHS}). A charcoal cryopump is used to evacuate or fill the circuit with $^{4}$He gas, opening and closing the switch respectively.

When the switch is evacuated, the only mechanism by which heat may pass through the switch is conduction through the thin-walled stainless steel tubes. Owing to the relatively low thermal conductivity of stainless steel at cryogenic temperatures and the small cross-sectional area of the tubes, the heat flow in this condition is minimal; this is the switch open state. When the cryopump is heated, the circuit is filled with gas and a convection current is established as long as the cold end is positioned higher with respect to gravity than the hot end (this may be achieved with copper heat straps). This provides a very effective mechanism for heat transfer and hence a high conductance through the switch in the closed position. The time required to fill and evacuate the tube (i.e., close or open the switch) is on the order of a few minutes.

Experimental measurements of the power through the switch in the open and closed states when the cold end is held at 4~K are shown in Figure~\ref{fig:precoolHScond1}, along with the modelled conductances\cite{May4}.

\subsection{1~K SORPTION COOLER}
\label{subsec:1Kcooler}

In order to cool the optics box from 4~K to 1~K and subsequently maintain 
temperature at 1~K during observations, a high power 1~K cooler\cite{may2016sorption} has been designed.  
The cooler is a $^{4}$He sorption refrigerator, as shown in Figure~\ref{fig:4He} 
below. 

\begin{figure}[h]
\centering
\begin{minipage}{.35\textwidth}
  \centering
  \includegraphics[width=0.85\linewidth]{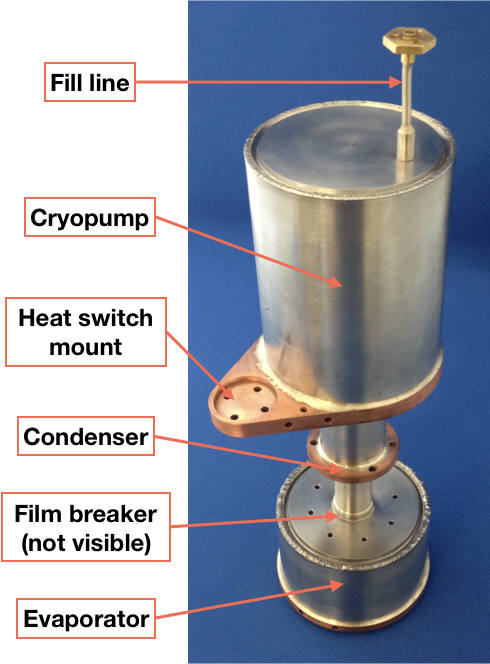}
  \caption{$^{4}$He 1~K cooler; the total height of the cooler is 300 mm}
  \label{fig:4He}
\end{minipage}%
\begin{minipage}{.6\textwidth}
  \centering
  \includegraphics[width=0.9\linewidth]{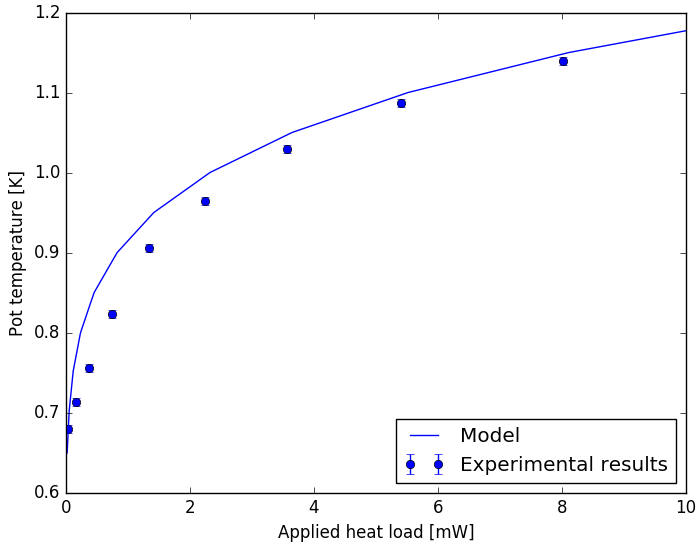}
  \caption{1~K cooler load curve theory and measurements}
  \label{figure_1-1}
\end{minipage}
\end{figure}

The cooler is charged to high pressure (70 bar) with $^{4}$He gas at room temperature; this is done from an external gas handling system through the fill line shown which is then permanently sealed.  
The cooler is mounted at the condenser flange to the 4~K stage and a heat switch  
also directly couples the charcoal cryopump to the 4~K stage.  The heat switch 
used is a convective design as used for precooling and described in Section 
\ref{subsec:precool}. When the cryostat is cooled, the 4~K stage cools the condenser flange 
directly as well as the cryopump (as the switch is closed at this point), causing the cryopump 
to adsorb all of the helium. In order to recycle the cooler, the switch is opened 
and the cryopump heated to 40~K causing the gas to desorb, condense in the 
condenser at 4~K and collect as liquid in the evaporator under the action of gravity. 

To run the cooler, the 
switch is then closed to cool the cryopump back to 4~K which causes it to pump 
on the liquid and cool the evaporator to $\sim$1~K (under typical loading).  The cooler will continue to 
operate until the liquid is completely adsorbed into the cryopump and the cycle 
is repeated. The cooler is therefore single shot; the design specification to 
support observation is a hold time at 1~K of 24 hours under a loading of 0.7 mW (see Section 
\ref{subsec:1Kbox}) with a recycling time of $\sim$2 hours. Figure~\ref{figure_1-1} below shows the measured and theoretical\cite{May3} load 
curve; the evaporator temperature under loading of 0.7 mW is 850 mK.

In this temperature regime, the liquid helium is a superfluid. A film will therefore 
tend to creep out of the evaporator and up the inside of the tube toward the condenser 
(i.e., in the direction of positive temperature gradient) under the action of the 
thermomechanical effect. This necessarily reduces the hold time of the system as 
the helium evaporates here without producing useful cooling; extensive efforts have been successfully devoted to developing a superfluid film 
breaker to minimise this effect\cite{May3}.

In order to initially cool the 1~K box from 4~K to 1~K,  60 J must be removed.  By using 
an isolating switch between the evaporator and the box, it is possible to cool 
the box to 1~K, recycle the cooler without significantly warming the box and 
then run the cooler again to maintain temperature during observation for $\sim$24 hours. When the cooler expires 
at the end of this observing period, the isolation switch is used again to avoid loading the box while 
the cooler is recycled. The switch design is described in 
detail in Section~\ref{subsec:isoswitch}.

\begin{figure}[h]
\centering
\begin{minipage}{.52\textwidth}
  \centering
  \includegraphics[width=\linewidth]{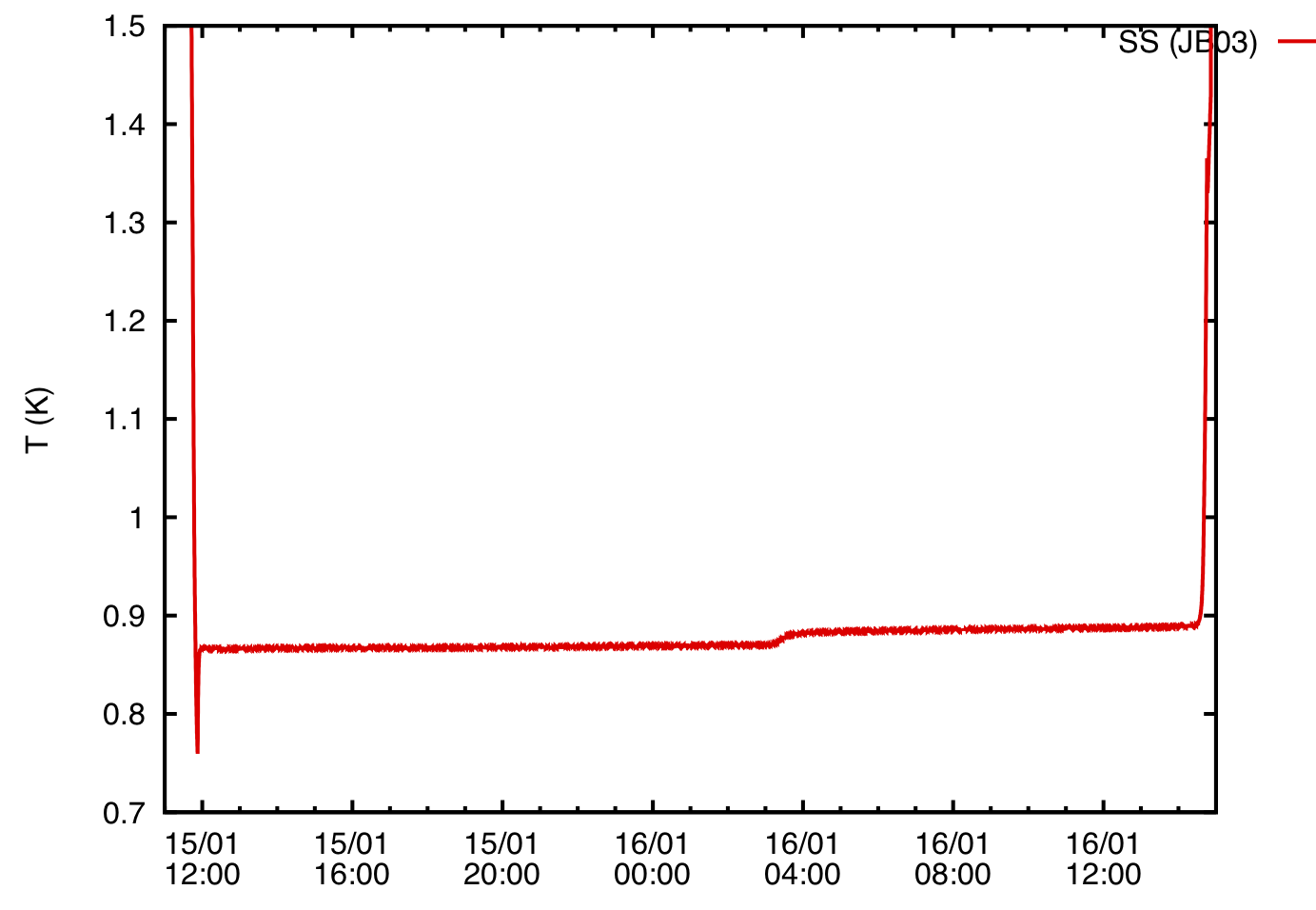}
  \caption{1 K cooler hold time for 0.7 mW applied load}
  \label{fig:holdtimeplot}
\end{minipage}%
\begin{minipage}{.48\textwidth}
  \centering
  \includegraphics[width=\linewidth]{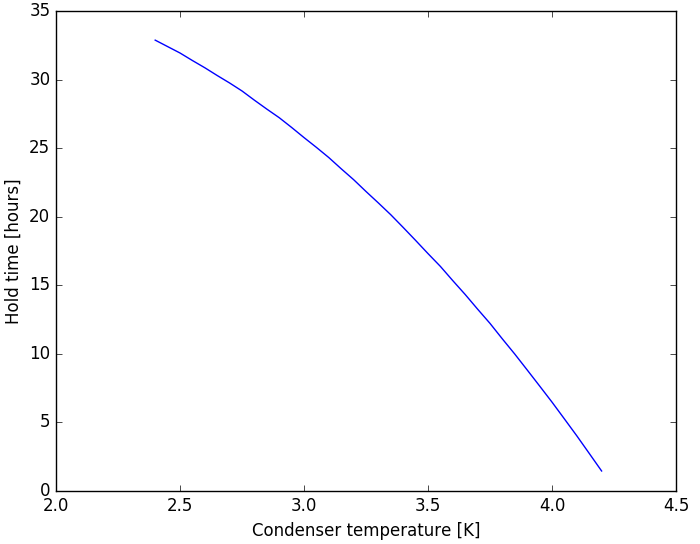}
  \caption{1 K cooler condensation efficiency}
  \label{fig:condeff}
\end{minipage}
\end{figure}

Experimental runs in a test cryostat under the design load of 0.7 mW showed a 
base temperature of 850 mK and a hold time of 26.5 hours, as 
shown in Figure~\ref{fig:holdtimeplot}. In this test, the condenser was cooled to 2.9 K, slightly colder than measured in the receiver cryostat cooldown
tests reported in Section~\ref{sec:4K}. Models of the condensation efficiency (and hence hold time)
show a dependency on the temperature of the condenser (see Figure 
\ref{fig:condeff}); according to this analysis,  a hold time of $\sim$24 hours will be achieved in the receiver cryostat.
The fridge, having been validated in Manchester has now been shipped to APC for integration with the 
receiver cryostat.

\subsection{ISOLATION HEAT SWITCH}
\label{subsec:isoswitch}

As described in Section~\ref{subsec:1Kcooler}, an isolation switch is required 
between the 1~K cooler cold head and the 1~K box. For this purpose, a $^{4}$He convective 
switch similar to that described in Section~\ref{subsec:precool} is used. 
However, in order to avoid the radiative loading associated with a 40~K cryopump 
when the switch is closed, a passive design (i.e., one with no cryopump) has instead been used.

The switch should be closed when the 1~K cooler is at base temperature (i.e., colder than the 1~K box) 
and open when the cooler is recycling (i.e., warmer than the 1~K box). As such, a switch of similar construction to the precooling switch (Figure \ref{fig:convHS}) is used with the 1~K cooler attached to the top of the switch. In this way, when the cooler is at base temperature the top of the switch will be colder than the bottom (attached to the 1~K box) and the switch couples the two stages strongly. Recycling the cooler will bring the top of the switch to a higher temperature than the bottom, killing the convection and opening the switch to isolate the box. In this way, no cryopump is required to actively control the switch which instead simply operates passively in analogy to a diode. In the receiver cryostat, copper support pieces are used to achieve the desired orientation with respect to gravity. 

\begin{figure}[!ht]
\centering
\begin{minipage}{.4\textwidth}
  \centering
  \includegraphics[width=.6\linewidth]{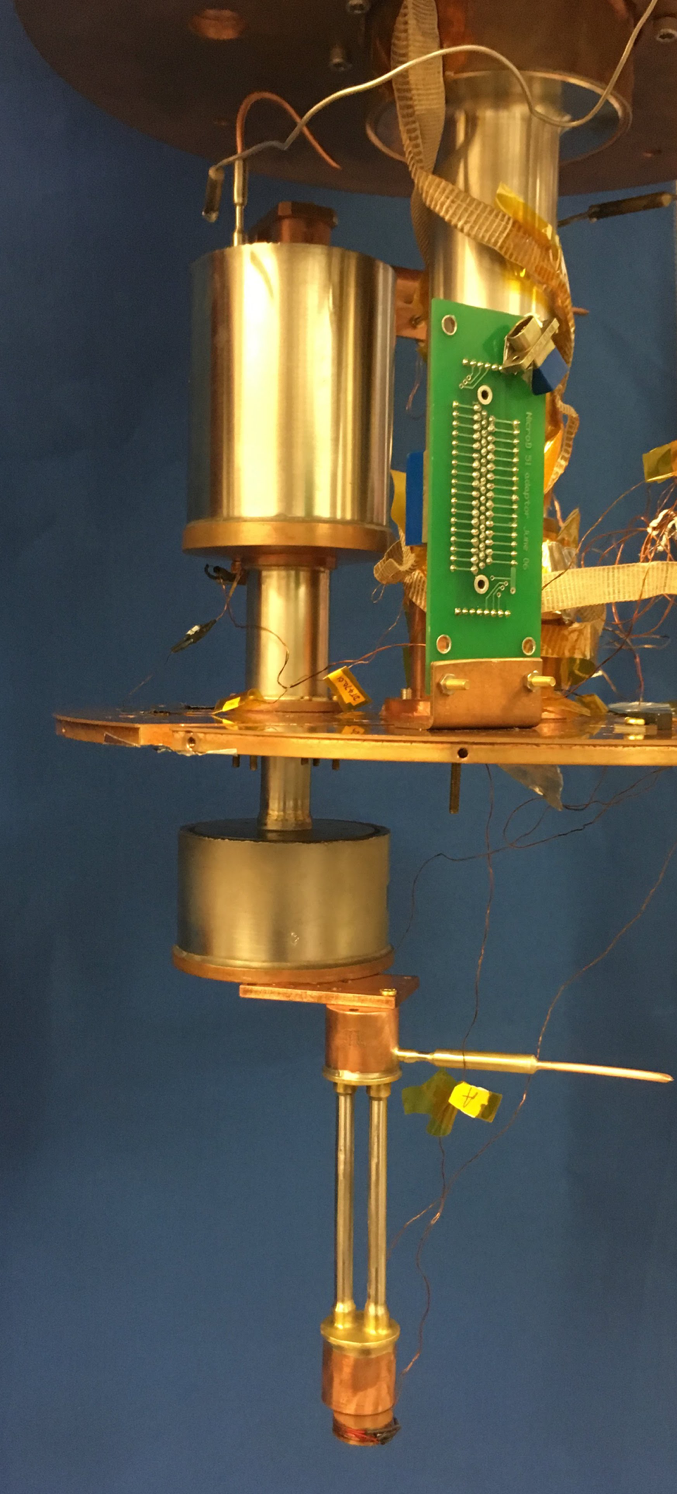}
  \caption{1~K cooler with passive isolation switch under test}
\label{fig:1Kcoolerpassive}
\end{minipage}%
\begin{minipage}{.6\textwidth}
  \centering
  \includegraphics[width=.9\linewidth]{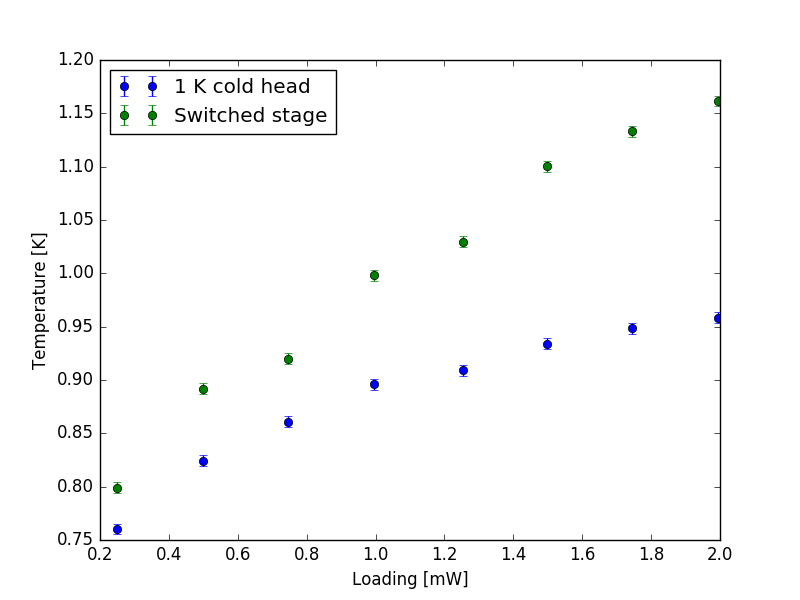}
\caption{Effective load curve of 1~K cooler with isolation switch}
\label{fig:1Kloadcurveeffective}
\end{minipage}
\end{figure}

The switch is shown under test in Figure~\ref{fig:1Kcoolerpassive}. As there is 
a small temperature step across the switch in the closed position, the use of 
the switch gives an ``effective'' load curve, as shown in Figure 
\ref{fig:1Kloadcurveeffective}. Under design load conditions, the temperature at the bottom of the switch is 910 mK (i.e., still meeting the design specification).

During testing, some residual conductance in the open state was observed below 
2.2 K as a result of superfluid film behaviour inside the tubes. Whilst the 
switch will be used as is for the TD, a $^{3}$He passive switch is currently 
being developed for use with the FI to improve cycling efficiency.

\section{350 MILLIKELVIN STAGE}
\label{sec:350mK}

Inside the 1~K box, a further cold stage is cooled to 350~mK to house the 
focal planes (see Figures~\ref{fig:instrument} and~\ref{fig:fullCAD}). The two orthogonal planes are tiled with
NbSi TES bolometers read out with a time domain multiplexing scheme as described 
in Section~\ref{sec:instrument}. The contributions to the thermal loading on 
this stage are given in Table~\ref{tab:350mKloading} below.

\begin{table}[h]
\centering
    \caption{350~mK stage steady-state thermal loading}
  \label{tab:350mKloading}
    \begin{tabular}{| l | l |}
  \hline			
  Source & Value \\
  \hline
  16 tie rods (epoxy/carbon-fibre) & 17 $\mu$W \\
  Instrumentation wires & 6 $\mu$W \\
  Thermal radiation through window & 1 $\mu$W \\
  Thermal radiation ($\varepsilon$=0.1) 1~K to 350~mK & $\ll$ 1 $\mu$W \\
  Total & 25 $\mu$W \\
\hline   
\end{tabular}
\end{table}

\begin{figure}[!ht]
\centering
\begin{minipage}{.45\textwidth}
  \centering
  \includegraphics[height=5cm]{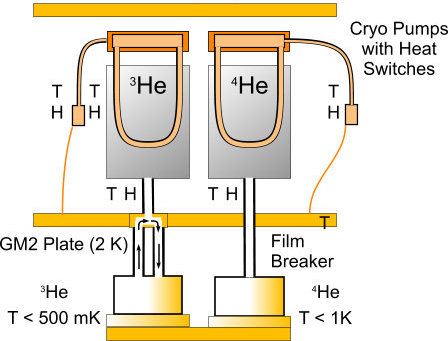}
\caption{Schematic of double-stage $^{3}$He/$^{4}$He 350~mK cooler}
\label{fig:ChaseSchem}
\end{minipage}%
\begin{minipage}{.5\textwidth}
  \centering
  \includegraphics[height=7cm]{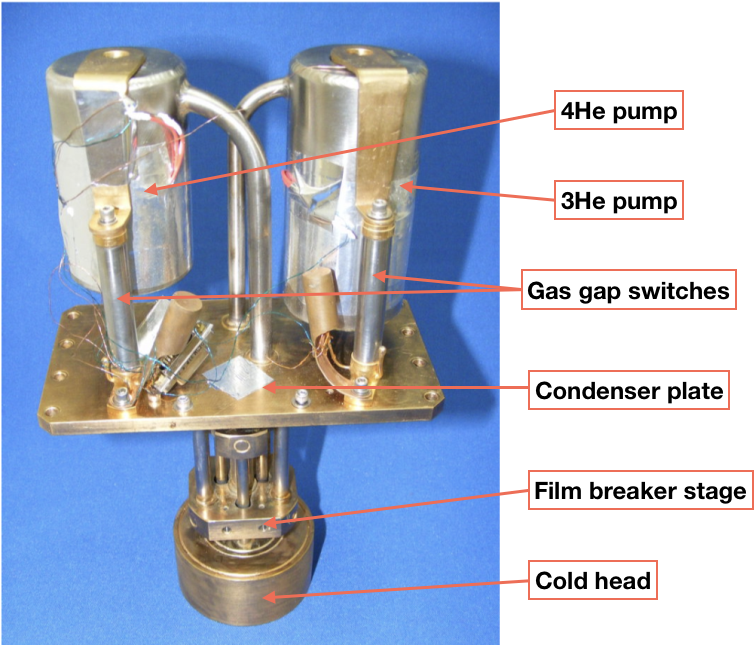}
\caption{Photograph of 350~mK cooler; the height of the cooler is 240 mm}
\label{fig:350mKcooler}
\end{minipage}
\end{figure}

A small double-stage $^{3}$He/$^{4}$He sorption cooler, commercially available from Chase Research Cryogenics\footnote{http://www.chasecryogenics.com} 
is used to cool the stage.  The cooler is mounted off the 4~K stage and operates 
in a similar manner to the 1~K cooler described in Section 
\ref{subsec:1Kcooler}. In this case however, the $^{4}$He stage is used only
to condense $^{3}$He in the second stage (sharing a common cold head), which cools to a lower temperature as 
a result of its different vapour pressure curve\cite{pobell2007matter}. Figure~\ref{fig:ChaseSchem} 
shows a schematic of the cooler. The cryopumps in this case are operated by gas 
gap heat switches\cite{piccirillo2018miniature}. A photograph of the cooler is 
shown in Figure~\ref{fig:350mKcooler}.

\begin{figure}[!ht]
\centering
\begin{minipage}{.45\textwidth}
  \centering
\includegraphics[width=\linewidth]{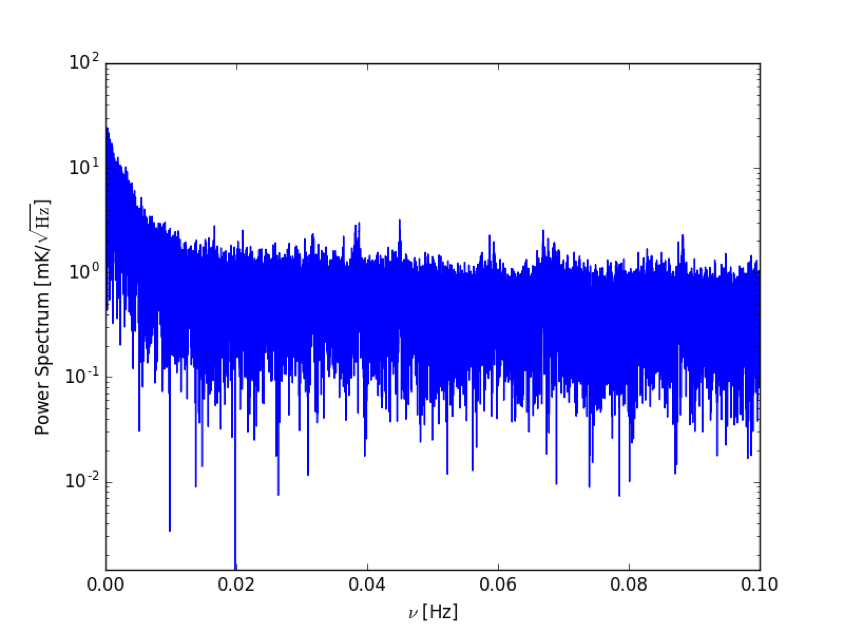}
\caption{Power spectral density for 350~mK cooler under design loading}
\label{fig:PSD}
\end{minipage}%
\begin{minipage}{.55\textwidth}
  \centering
  \includegraphics[width=\linewidth]{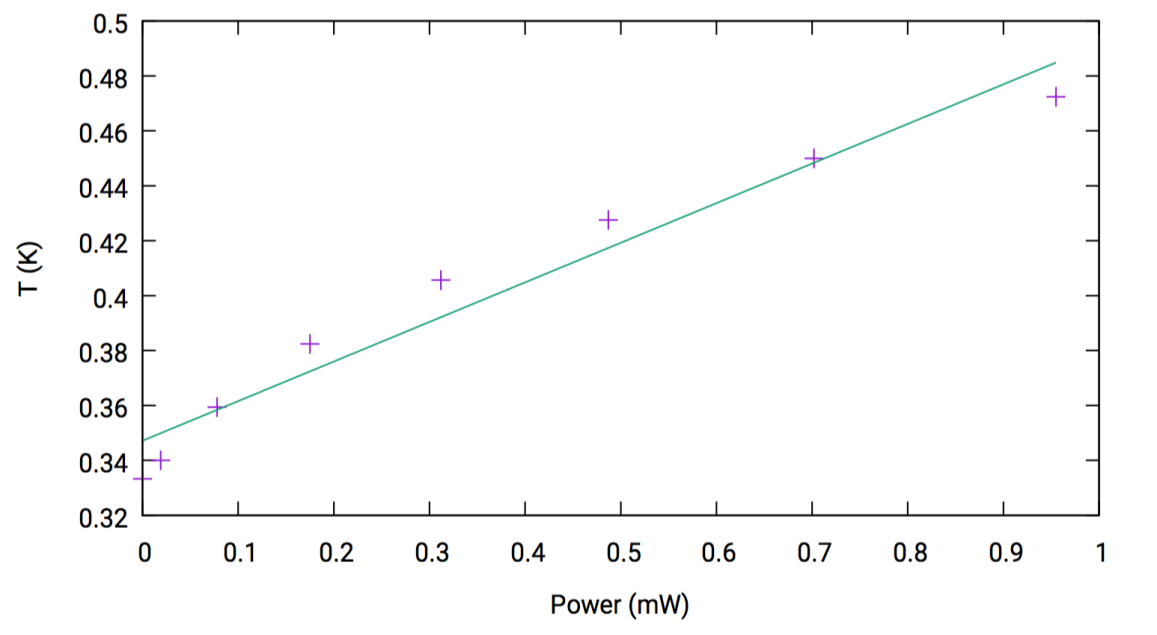}
\caption{350~mK cooler load curve measurements} 
\label{fig:350mKloadcurve}
\end{minipage}
\end{figure}

The cooler was tested under the design load and showed a hold time of 90 hours and a base temperature of 336 
mK. A full load curve is shown in Figure~\ref{fig:350mKloadcurve}. Calculation of 
the power spectral density, as shown in Figure~\ref{fig:PSD}, showed excellent temperature 
stability, with negligible long term drift observed.

\section{SUMMARY}

In order to reach the sensitivity required for B-mode polarization observations, the QUBIC experiment requires an exceptionally large and complex cryostat, housing a large number of novel and highly optimized optical elements on stages at 40~K, 4~K, 1~K and 350~mK. Design and development of the cryostat has been particularly challenging. 

Extensive analysis, laboratory testing and design efforts have been required to develop both the cryostat and the necessary subsystems for the sub-4~K stages. A novel precooling strategy based on a combination of heat switches has been developed to support reasonably quick turnaround time for testing and deployment. A high-power $^{4}$He 1~K cooler with a novel superfluid film breaker for extended hold time has been developed, along with an isolation switch to improve recycling efficiency. A small commercial double stage cooler has been extensively tested and characterised for use in cooling the TES focal plane array to 350~mK.

The cryostat vacuum chamber, 40~K and 4~K stages have been succesfully demonstrated in Roma. The cold optical components and sub-4~K cryogenics systems have also been succesfully tested at a number of other institutions. All systems have now been shipped to APC for integration into the TD during Summer 2018, representing a crucial milestone toward deployment of the first 
module\cite{osullivan}.

\acknowledgments

The authors would like to acknowledge the support of INFN in Italy and IN2P3 in France. The APC team acknowledges the support of the UnivEarthS Labex program at Sorbonne Paris Cit\'{e} (ANR-10-LABX-0023 and ANR-11-IDEX-0005-02). 
The University of Manchester team would like to acknowledge the support of STFC grant 
ST/L000768/1 and the assistance of the School of Physics and Astronomy workshop in manufacture of 
many of the sub-4~K systems detailed in this paper. AJM is supported by an STFC PhD studentship and his attendance is supported by the Institute of Physics Low Temperature Group.

\bibliography{report} 
\bibliographystyle{spiebib} 

\end{document}